\documentclass[epj]{svjour}
\usepackage[latin1]{inputenc}
\usepackage{graphicx}
\usepackage{amsmath,amssymb}
\usepackage{cite}
\graphicspath{{./img/}}

\renewcommand{\d}{\mathrm{d}}
\newcommand{\e}{\mathrm{e}}
\newcommand{\tr}{\operatorname{tr}}
\renewcommand{\Re}{\operatorname{Re}}

\renewcommand{\i}{\mathrm{i}}
\newcommand{\void}[1]{}

\renewcommand{\L}{\mathrm{L}}
\newcommand{\R}{\mathrm{R}}
\newcommand{\bra}[1]{\langle #1|}
\newcommand{\ket}[1]{|#1 \rangle}
\newcommand{\exponent}[1]{\ensuremath{\mathrm e^{#1}}}
\begin{document}
\title{Coulomb repulsion effects in driven electron transport}
\author{Franz J. Kaiser \and Peter H\"anggi \and Sigmund Kohler}
\institute{Institut f\"ur Physik, Universit\"at Augsburg,
Universit\"atsstra\ss e~1, D-86135 Augsburg, Germany}

\date{19 April 2007}
\abstract{
We study numerically the influence of strong Coulomb repulsion on the
current through molecular wires that are driven by external electromagnetic
fields.  The molecule is described by a tight-binding model whose first
and last site is coupled to a respective lead.  The leads are
eliminated within a perturbation theory yielding a master equation for
the wire.  The decomposition into a Floquet basis enables an efficient
treatment of the driving field.
For the electronic excitations in bridged molecular wires, we find that
strong Coulomb repulsion significantly sharpens resonance peaks which broaden
again with increasing temperature.
By contrast, it has only a small influence on effects
like non-adiabatic electron pumping and coherent current suppression.
}

\maketitle

\section{Introduction}

Recent experiments on the conductance of single organic molecules
opened a new direction in mesoscopic transport \cite{Reed1997a,
Cui2001a, Reichert2002a}.  Of particular interest is thereby the
influence of electronic and vibronic excitations of the molecules
which leave their fingerprints in the resulting current-voltage
characteristics.  At low temperatures, these effects become more
pronounced \cite{Reichert2003a}.
Much of our knowledge about excitations of molecules is
based on spectroscopy, i.e.\ the optical response to light.  In the
context of molecular conduction, it has been proposed to study as well
the signatures of such excitations in the transport quantifiers like
the current \cite{Kohler2002a, Keller2002a, Tikhonov2002b} and its
fluctuations \cite{Kohler2005a}.
Such experiments are at present attempted, but clearcut evidence for
the proposed effects is still missing because the irradiation also
causes unwanted thermal effects in the contacts, which in today's
setups seem to dominate.  One possibility to protect the contacts
against the radiation is using the evanescent light at the tip of a
near-field optical microscope.

Coupled quantum dots represent a setup with properties similar to
those of molecular wires \cite{Blick1996a, vanderWiel2003a}, albeit at
different length and energy scales.  As compared to molecular wires,
they are more stable and tunable, but have the disadvantage that
only a few dots can be coupled coherently.  The transport properties
of these ``artificial molecules'' can be significantly modified by
microwaves \cite{vanderWiel2003a}.  It has for example been
demonstrated experimentally that resonant excitations between the
levels of double quantum dots result in the so-called photon-assisted
transport, i.e.\ a significant enhancement of the dc current
\cite{Oosterkamp1998a, vanderWiel1999a}.

In addition to photon-assisted transport \cite{Platero2004a}, other
less intuitive phenomena have been predicted in this context.
For example the so-called coherent suppression of
tunnelling in a double-well potential due to a time-dependent dipole
force can also be found in ac driven transport through coupled
quantum dots: For characteristic ratios between the amplitude and the
frequency of the driving, it has been predicted that the dc current
\cite{Lehmann2003a, Grossmann2004a, Welack2006a} and the shot noise
\cite{Camalet2003a} will be suppressed.
A further prominent effect is adiabatic electron pumping, which is
the generation of a dc current by means of a periodic variation of the
conductor parameters in the absence of any net bias \cite{Brouwer1998a,
Altshuler1999a, Switkes1999a, Wang2002a}.  It has been proposed
\cite{Stafford1996a, Wagner1999a} and experimentally demonstrated
\cite{Oosterkamp1998a, vanderWiel2003a} that pumping is more effective
at internal resonances, i.e., beyond the adiabatic limit where, in
addition, the pump current possesses a surprisingly low noise level
\cite{Strass2005b}.

Periodically time-dependent quantum systems can be described very
efficiently within a Floquet theory which originally has been derived
for driven closed quantum systems \cite{Shirley1965a} and later been
generalised to dissipative quantum systems \cite{Kohler1997a, Grifoni1998a}.
Furthermore, it is possible to derive Floquet theories for the
description of transport through mesoscopic conductors which are
connected to external leads.  For cases in which electron-electron
interactions do not play any role, one can derive a Floquet scattering
theory that provides exact expressions for the current
\cite{Moskalets2002a, Camalet2003a} and its noise \cite{Camalet2003a,
Kohler2005a}.
Treating the coupling between the conductor and the leads perturbatively, one
can obtain a master equation for the reduced density operator of the wire.  This
enables a rather efficient treatment of time-dependent transport after
decomposing the wire density operator into a Floquet basis.  Then it is possible
to study relatively large driven conductors \cite{Lehmann2002b} and to include
also electron-electron \cite{Bruder1994a} and electron-phonon interactions
\cite{Lehmann2004a}. If the time-dependent field consists of one or a
few laser pulses, it is possible to obtain the density operator by
propagating the Liouville-von Neumann equation
\cite{Kleinekathofer2006a}.

In this work, we derive in Sections \ref{sec:model} and
\ref{sec:theory} a Floquet master equation formalism that captures
situations in which strong Coulomb repulsion restricts the excess
charge residing on the conductor to a single electron.  For later
reference, we adapt in Section \ref{sec:other} our approach to the
case of spinless electrons and non-interacting electrons.  We then
investigate in Section \ref{sec:applications} the role of strong
Coulomb repulsion for photon-assisted transport through bridged molecular
wires, non-adiabatic electron pumping, and coherent current
suppression.

\section{The wire-lead model}
\label{sec:model}

\begin{figure}[tb]
\centerline{\includegraphics{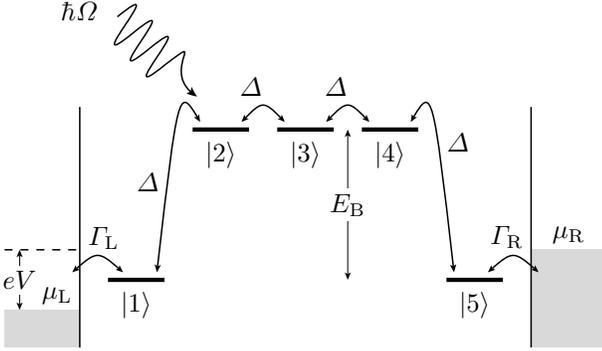}}
\caption{Bridged molecular wire model consisting of $N=5$ sites with internal
tunnelling matrix elements $\Delta$ and effective wire-lead coupling
strengths $\Gamma_\mathrm{L/R}$.}
\label{fig:wire_model}
\end{figure}
The system of the driven wire, the leads, and the coupling between the molecule
and the leads, as sketched in Fig.~\ref{fig:wire_model}, is described by
the Hamiltonian
    \begin{equation}
    \label{eq:full_hamiltonian}
    \mathcal{H}(t) = \mathcal{H}_{\text{wire}}(t) + \mathcal{H}_{\text{leads}} +
    \mathcal{H}_{\text{wire-leads}}.
    \end{equation}
The wire is modelled within a tight-binding description by the molecular
orbitals $\ket{n}$, $n=1,\ldots,N$, so that
    \begin{equation}
    \label{eq:wire_hamiltonian}
    \mathcal{H}_{\text{wire}}(t) = \sum_{n,n',s} H_{nn'}(t)\,
    c_{n\vphantom{'}s}^\dagger c^{\vphantom{\dagger}}_{n's}
    + \frac{U}{2} N_\mathrm{wire}(N_\mathrm{wire}-1) ,
    \end{equation}
where $c_{ns}^\dagger$ ($c_{ns}$)
creates (annihilates) an electron with spin $s$ at site $\ket{n}$ and
$[c_{ns}, c^{\dagger}_{n's'}]_+= \delta_{n,n'} \delta_{s,s'} $.
The influence of a driving field entails a $\mathcal{T}$-periodic
time-dependence on the single-particle Hamiltonian $H_{nn'}(t) =
H_{nn'}(t+\mathcal{T})$.  The last term in
Eq.~\eqref{eq:wire_hamiltonian} captures the electron-electron
interaction within a capacitor model and the operator $N_\mathrm{wire}
= \sum_{n,s} c_{ns}^\dagger c_{ns}$ describes the number of excess
electrons residing on the molecule.  Below we shall assume that $U$ is
so large that only states with zero or one excess electron play a
role.

The first and the last site of the molecule, $\ket{1}$ and $\ket{N}$,
couple via the tunnelling Hamiltonian
    \begin{equation}
    \label{eq:tunnel_hamiltonian}
    \mathcal{H}_{\text{wire-leads}} =  \sum_{q,s} (V_{\L q}\,
    c^{\dagger}_{\L qs}
    c_{1s}^{\vphantom{\dagger}} + V_{\R q}\, c^{\dagger}_{\R qs}
    c_{Ns}^{\vphantom{\dagger}}) + \text{H.c.}
    \end{equation}
to the respective lead.  The operator $c^{\dagger}_{\L qs}$
($c^{\dagger}_{\R qs}$) creates an electron in the left (right) lead
in the state $|\L qs\rangle$ which is orthogonal to all wire states.
It will turn out that the influence of the tunnelling Hamiltonian is
fully characterised by the spectral density
\begin{equation}
  \label{spectral.density}
  \Gamma_\ell(\epsilon) =
         2\pi\sum_q |V_{\ell q}|^2 \delta(\epsilon-\epsilon_q)
\end{equation}
which becomes a continuous function of the energy $\epsilon$ if the
lead states are dense.  If all relevant lead states are located in the
centre of the conduction band, the energy-dependence of the spectral
densities is not relevant so that they can be replaced by a constant,
$\Gamma_\ell(\epsilon) = \Gamma_\ell$.  This defines the so-called
wide-band limit.

The leads are modelled as ideal Fermi gases
    \begin{equation}
    \label{eq:lead_hamiltonian}
    \mathcal{H}_{\text{leads}} =  \sum_{q,s} \big(
    \epsilon^{\vphantom{\dagger}}_{q}
    \, c^{\dagger}_{\L qs} c^{\vphantom{\dagger}}_{\L qs}+
    \epsilon^{\vphantom{\dagger}}_{q}
    \, c^{\dagger}_{\R qs} c^{\vphantom{\dagger}}_{\R qs}
    \big) ,
    \end{equation}
which are initially at thermal equilibrium with the chemical potential
$\mu_{\L/\R}$ and, thus, are described by the density operator
\begin{equation}
    \rho_{\text{leads,eq}} \propto
    \exp{[-(\mathcal{H}_{\text{leads}}-\mu_\L N_\L-\mu_\R
    N_\R)/k_\text{B}T]},
\end{equation}
where $N_\ell = \sum_q c_{q\ell}^\dagger c_{q\ell}$ denotes the
electron number in lead $\ell = \L,\R$.  Then all lead properties can be
expressed in terms of the expectation value
\begin{equation}
    \langle c_{\ell qs}^\dagger c_{\ell'q's'}^{\vphantom{\dagger}}\rangle =
    \delta_{\ell\ell'} \delta_{qq'} \delta_{ss'} f_{\ell}(\epsilon_{q}),
\end{equation}
where $f_\ell(\epsilon) =
(1+\exponent{(\epsilon-\mu_\ell)/k_\text{B}T})^{-1}$ denotes the Fermi
function.
Since a typical metal screens all electric fields with a frequency below the
plasma frequency, we assume that the bulk properties of the leads are
not affected by the laser irradiation.

\section{Master equation approach for strong Coulomb repulsion}
\label{sec:theory}

\subsection{Perturbation theory}

Most master equation approaches to electron transport
\cite{Nazarov1993a, Bruder1994a, Gurvitz1996a, Lehmann2004a,
Kleinekathofer2006a} are based
on a perturbative treatment of the wire-lead Hamiltonian
$H_\text{wire-leads}$.  Within these approaches, it is possible to
include within a Floquet theory the response to a time-periodic field
exactly \cite{Brune1997a, Kohler2005a}.  The derivation of a master
equation starts from the Liouville-von Neumann equation $\i\hbar \dot
\rho(t)=[H(t),\rho(t)]$ for the total density operator $\rho(t)$ for
which one obtains by standard techniques the approximate equation of
motion
\begin{align}
\label{mastereq-gen}
  \dot\rho(t)
  = & -\frac{\i}{\hbar}[H_\mathrm{wire}(t)+H_\mathrm{leads},\rho(t)] \\
    & -\frac{1}{\hbar^2}\int_0^\infty \d\tau
      [H_\text{wire-leads},[\widetilde
      H_\text{wire-leads}(t-\tau,t),\rho(t)]] . \nonumber
\end{align}
Here the first term corresponds to the coherent dynamics of both the wire
electrons and the lead electrons, while the second term describes
resonant electron tunnelling between the leads and the wire.
The tilde denotes operators in the interaction picture with respect to
the molecule and the lead Hamiltonian without the molecule-lead
coupling, $\widetilde X(t,t')=U_0^\dagger(t,t')\,X\,U_0(t,t')$, where $U_0$
is the propagator without the coupling.  The net (incoming minus
outgoing) electrical current through the left contact is given by
minus the time-derivative of the electron number in the left lead
multiplied by the electron charge $-e$.  From Eq.~(\ref{mastereq-gen})
follows for the current in the wide-band limit the expression
\begin{align}
  I_\L(t)
  ={}&  e \mathop{\text{tr}}[\dot \rho(t) N_\L] \nonumber
\\
  ={}&
   -e\frac{\Gamma_\L}{\pi\hbar}\mathop{\text{Re}}\int_0^\infty
   \d\tau \int\d\epsilon\, \e^{\i \epsilon \tau/\hbar}
   \label{current-general} \\
   & \times
    \big\{ \langle c^\dagger_1 \tilde{c}_1^{\vphantom{\dagger}}
    (t,t-\tau)\rangle 
    \bar{f}_\L (\epsilon) -
    \langle \tilde{c}_1^{\vphantom{\dagger}}(t,t-\tau)c_1^\dagger
    \rangle 
    f_\L(\epsilon)\big\} ,
\nonumber
\end{align}
where $\bar f_\ell = 1-f_\ell$ and $\langle\cdots\rangle =
\mathop{\text{tr}_\text{wire}}\rho_\text{wire}\cdots$ denotes the
average with respect to the wire density operator, which still has to
be determined.
We emphasise that the expectation values in
Eq.~\eqref{current-general} depend directly on the dynamics of the
isolated wire and are thus influenced by the driving.

\subsection{Floquet theory}

An important feature of the current formula \eqref{current-general} is its
dependence on solely the wire operators while all lead operators have
been eliminated.  Therefore it is sufficient to consider the reduced
density operator $\rho_\text{wire} =
\mathop{\text{tr}_\text{leads}}\rho$ of the wire electrons.  However,
$\rho_\text{wire}$ still obeys a time-dependent master equation whose
direct solution requires quite some effort, in particular if one is
interested in the behaviour at long times.
This effort can be reduced significantly by exploiting the fact that
the master equation \eqref{mastereq-gen} inherited from the total
Hamiltonian $\mathcal{H}(t)$ a $\mathcal{T}$-periodic
time-dependence, which opens the way for a Floquet treatment.

\subsubsection{Fermionic Floquet operators}

One possibility for such a treatment is to use the Floquet states of
the central system, i.e.\ the driven wire, as a basis.  Thereby we
also use the fact that in the wire Hamiltonian
\eqref{eq:wire_hamiltonian}, the single-particle contribution commutes with
the interaction term and, thus, these two Hamiltonians possess a complete set
of common eigenstates.  Here we start by diagonalising the first part
of the Hamiltonian which describes the single-particle dynamics
determined by the time-periodic matrix elements $H_{nn'}(t)$.
According to the Floquet theorem, the corresponding (single particle)
Schr\"odinger equation possesses a complete solution of the form
\begin{equation}
    \ket{\Psi_{\alpha}(t)}
    =
    \exponent{\i \epsilon_{\alpha} t / \hbar} 
    \ket{\varphi_{\alpha}(t)},
\end{equation}
with the so-called quasienergies $\epsilon_\alpha$ and the
$\mathcal{T}$-periodic Floquet states
\begin{equation}
    \ket{\varphi_{\alpha}(t)} =  \sum_{k} \exponent{- \i k \Omega t}
    \ket{\varphi_{\alpha,k}}.
\end{equation}
The Floquet states and eigenenergies are obtained by solving the
eigenvalue problem
\begin{equation}
    \label{eq:quasienergy}
    \Big( \sum_{n,n'} \ket{n} H_{nn'}(t) \bra{n'}  -\i \hbar
    \frac{\text{d}}{\text{d}t} \Big) \ket{\varphi_{\alpha}(t)}
  = \epsilon_{\alpha} \ket{\varphi_{\alpha}(t)} ,
\end{equation}
whose solution allows one to construct via Slater determinants
many-particle Floquet states.
In analogy to the quasimomenta in Bloch theory for spatially periodic
potentials, the quasienergies $\epsilon_{\alpha}$ come in classes
\begin{equation}
    \epsilon_{\alpha,k}=\epsilon_\alpha + k\hbar \Omega,
    \quad k \in  \mathbb{Z},
\end{equation}
of which all members represent the same physical solution of the Schr\"odinger
equation. Thus we can restrict ourselves to states within one Brillouin
zone like for example $0 \leq \epsilon_\alpha < \hbar \Omega$.

For the computation of the current it is convenient to have an
explicit expression for the interaction picture representation of the
wire operators.  It can be obtained from the (fermionic) Floquet creation
and annihilation operators \cite{Lehmann2003b} defined via the
transformation
\begin{equation}
    \label{c.alpha}
    c_{\alpha s} (t) = \sum_n \bra{\varphi_{\alpha}(t)} n\rangle c_{ns} .
\end{equation}
The inverse transformation
\begin{equation}
    \label{eq:floquet_operator}
    c_{ns} = \sum_\alpha \langle n \ket{\varphi_{\alpha}(t)} c_{\alpha s}(t)
\end{equation}
follows from the mutual orthogonality and the completeness of the
Floquet states at equal times \cite{Grifoni1998a}.  Note that the
right-hand side of Eq.~\eqref{eq:floquet_operator} becomes time
independent after the summation.

The Floquet annihilation operator \eqref{c.alpha} has the interaction
picture representation
\begin{eqnarray}
  \tilde{c}_{\alpha s}(t,t')
  &=& U^\dagger_0 (t,t')\, c_{\alpha s}(t)\,  U^{\vphantom{\dagger}}_0 (t,t')
    \\
  &=& \exponent{- \i(\epsilon_{\alpha}+U N_\text{wire}) (t-t') / \hbar}
    c_{\alpha s}(t') ,
  \label{c.alpha.interaction}
\end{eqnarray}
with the important feature that the time difference $t-t'$ enters
only via the exponential prefactor.  This will allow us to evaluate
the $\tau$-integration of the master equation \eqref{mastereq-gen}
after a Floquet decomposition.
Relation \eqref{c.alpha.interaction} can easily be shown by computing
the time derivative with respect to $t$ which by use of the Floquet
equation \eqref{eq:quasienergy} becomes
\begin{equation}
\label{c-t-t'}
   \frac{\d}{\d t} \tilde c_{\alpha s}(t,t')
 = -\frac{\i}{\hbar} (\epsilon_\alpha + UN_\text{wire})\,
   \tilde c_{\alpha s}(t,t') .
\end{equation}
Together with the initial condition $\tilde c_\alpha(t',t') =
c_\alpha(t')$ follows relation \eqref{c.alpha.interaction}.
Note that the time evolution induced by $\mathcal{H}_\text{wire}(t)$
conserves the number of electrons on the wire.

\subsubsection{Master equation and current formula}

In order to make use of the Floquet ansatz, we decompose the master
equation \eqref{mastereq-gen} and the
current formula \eqref{current-general} into the Floquet basis derived
in the last subsection.  For that purpose we use the fact that
we are finally interested in the current at asymptotically large times
in the limit of a large interaction $U$.
The latter has the consequence that only wire states with at most one
excess electron play a role, so that the density operator
$\rho_\text{wire}$ can be decomposed into the $2N+1$ dimensional basis
$\{ |0\rangle, c_{\alpha s}^\dagger(t)\,|0\rangle \}$, where
$|0\rangle$ denotes the wire state in the absence of excess electrons
and $s=\uparrow,\downarrow$.
Moreover, it can be shown that at large times, the density operator of
the wire becomes diagonal in the electron number $N_\text{wire}$.
Therefore a proper ansatz reads
\begin{equation}
  \label{eq:decomposition}
  \rho_{\text{wire}}(t)
  = |0\rangle \rho_{00}(t) \langle 0|
    + \sum_{\alpha,\beta,s,s'} c^\dagger_{\alpha s}
    \ket{0}\rho_{\alpha s,\beta s'}(t)\bra{0}c_{\beta s'} .
\end{equation}
Note that we keep terms with $\alpha\neq\beta$, which means that we
work beyond a rotating-wave approximation.  Indeed in a
non-equilibrium situation, the off-diagonal density matrix elements
$\rho_{\alpha\beta}$ will not vanish and neglecting them might lead to
artefacts \cite{Novotny2002a, Kohler2005a}. In the context of
molecular wires, such a treatment of strong Coulomb repulsion by a
restriction to at most one excess electron has recently also been
applied to incoherent \cite{Petrov2001a, Petrov2002a,
Lehmann2002a} as well as to coherent transport \cite{Kaiser2006a}.

By inserting the decomposition \eqref{eq:decomposition}
into the master equation \eqref{mastereq-gen}, we obtain an equation
of motion for the matrix elements $\rho_{\alpha s,\beta s'}
= \langle0|c_{\alpha s}\rho_\text{wire} c^\dagger_{\beta s'}|0\rangle$.
We evaluate the trace over the lead states and compute the
matrix element $\langle 0|c_{\alpha s}(t)\ldots c_{\beta s'}^\dagger(t)|0\rangle$.
Thereby we neglect the two-particle terms which are of the structure
$c_{\alpha s}^\dagger c_{\beta s}^\dagger |0\rangle\langle 0| c_{\beta s}
c_{\alpha s}$.  Formally, these terms drop out in the limit of strong
Coulomb repulsion because they are accompanied by a rapidly oscillating
phase factor $\exp({-\i
UN_\text{wire}\tau}/\hbar)$.  Then the $\tau$-integra\-tion results in a
factor $f_\L(\epsilon_{\alpha,k}+U)$ which vanishes in the limit of
large $U$.  Since the total Hamiltonian
\eqref{eq:full_hamiltonian} is diagonal in the spin index $s$, we find
that the density matrix elements $\rho_{\alpha s, \beta s'}$ are
spin-independent as well so that after a transient stage
\begin{equation}
  \rho_{\alpha {\uparrow},\beta {\uparrow}}(t)
  =\rho_{\alpha {\downarrow},\beta {\downarrow}}(t)
  \equiv \rho_{\alpha\beta}(t)
\end{equation}
and $\rho_{\alpha {\uparrow},\beta {\downarrow}} =0$.
Moreover, at large times, the density operator \eqref{eq:decomposition} will
acquire the time periodicity of the driving field \cite{Kohler2005a}
and, thus, can be decomposed into the Fourier series
\begin{equation}
\label{rho-fourier}
\rho_{\alpha\beta}(t) = \sum_k \e^{-\i k\Omega t} \rho_{\alpha\beta,k}
\end{equation}
and $\rho_{00}(t)$ accordingly.

After some algebra, we arrive at a set of $N^2$ coupled equations of motion for
$\rho_{\alpha\beta}(t)$ which in Fourier representation read
\begin{equation}
\label{eq:masterfourier}
    \begin{alignedat}{1}
    \i( & \epsilon_{\alpha}-\epsilon_{\beta}-k\hbar\Omega)
    \rho_{\alpha \beta,k}
    \\
   =&\frac{\Gamma_\L }{2}\sum_{k',k''}\,
    \langle \varphi_{\alpha,k'+k''} | 1 \rangle 
    \langle 1 | \varphi_{\beta, k+k''} \rangle
    \rho_{00,k'}\\
    &\hphantom{\frac{\Gamma_\L }{2}\sum_{k',k''}\,\langle
    \varphi_{\alpha,k'+k''} | 1 \rangle}
    \times \big( f_\L(\epsilon_{\alpha,k'+k''})
    +f_\L(\epsilon_{\beta,k+k''}) \big)
    \\
    &-\frac{\Gamma_\L }{2}\sum_{\alpha',k',k''}
    \langle \varphi_{\alpha,k'+k''} | 1 \rangle
    \langle 1 | \varphi_{\alpha',k+k''} \rangle
    \rho_{\alpha'\beta, k'}\,
    \bar{f}_\L (\epsilon_{\alpha',k+k''})
    \\
    &-\frac{\Gamma_\L }{2}\sum_{\beta',k',k''}
    \langle \varphi_{\beta',k'+k''} | 1 \rangle
    \langle 1 | \varphi_{\beta,k+k''} \rangle
    \rho_{\alpha\beta', k'}\,
    \bar{f}_\L (\epsilon_{\beta',k'+k''})
    \\
    &+\text{same terms with the replacement}\:
    1, \L \rightarrow N, \R.
    \end{alignedat}
\end{equation}
In order to solve these equations, we have to eliminate
$\rho_{00,k}$ which is most conveniently done by inserting the Fourier
representation of the normalisation condition
\begin{equation}
\label{trace}
\tr\rho_\text{wire}(t)
= \rho_{00}(t) + 2\sum_{\alpha} \rho_{\alpha\alpha}(t) = 1 .
\end{equation}

In order to obtain for the current an expression that is
consistent with the restriction to one excess electron, we compute the
expectation values in the current formula \eqref{current-general} with
the reduced density operator \eqref{eq:decomposition} and insert the
Floquet representation \eqref{eq:floquet_operator} of the wire
operators.  Performing an average over the driving period, we obtain
for the dc current the expression
\begin{equation}
\label{eq:dc-current}
\begin{split}
    I = \frac{2e \Gamma_\L}{\hbar} \Re \sum_{\alpha,k}\Big(
    &
    \sum_{\beta,k'}
    \langle \varphi_{\beta,k'+k} | 1 \rangle
    \langle 1 | \varphi_{\alpha,k} \rangle
    \rho_{\alpha \beta, k'}
    \bar{f}_\L(\epsilon_{\alpha,k})
    \\ - &
    \sum_{k'}
    \langle \varphi_{\alpha,k'+k} | 1 \rangle
    \langle 1 | \varphi_{\alpha,k} \rangle 
    \rho_{00,k'}
    f_\L(\epsilon_{\alpha,k})
    \Big).
\end{split}
\end{equation}
Physically, the second contribution of the current formula
\eqref{eq:dc-current} describes the tunnelling of an electron from the
left lead to the wire and, thus, is proportional to $\rho_{00} f_\L$
which denotes the probability that a lead state is occupied while the
wire is empty.  The first terms corresponds to the reversed process
namely the tunnelling on an electron from site $|1\rangle$ to the left
lead.

The results of this section allow us the numerical computation of the dc
current through a driven conductor in the
the following way: First, we solve the quasienergy equation
\eqref{eq:quasienergy} which provides the coefficients
$\langle\varphi_{\alpha,k}|n\rangle$.  Next, we solve the master equation
\eqref{eq:masterfourier} and insert the solution into the current
formula \eqref{eq:dc-current}.

\section{Separating interaction and spin}
\label{sec:other}

In order to determine the role of a strong interaction, we
shall compare below our results to the non-interacting case.
Moreover, a particular consequence of strong Coulomb repulsion is the
mutual blocking of different spin channels.  This motivates us to also
compare to the case of spinless electrons which is physically realised
by spin polarisation.
In this section, we adapt our master equation approach to these
situations.

\subsection{Spinless electrons}

In order to describe spinless electrons, we drop in the initial
Hamiltonian all spin indices.  Physically, this limit is realised by
a sufficiently strong magnetic field that polarises all
electrons contributing to the transport.
By the same calculation as in Section \ref{sec:theory}, we then obtain for
the current also the expression \eqref{eq:dc-current} but without the
prefactor $2$.  The factor $2$ is also no longer present in the normalisation
condition \eqref{trace} which now reads
\begin{equation}
\label{trace-nospin}
\tr\rho_\text{wire}(t)
= \rho_{00}(t) + \sum_\alpha \rho_{\alpha\alpha}(t) = 1 .
\end{equation}

\subsection{Non-interacting electrons}

In the absence of interactions, $U=0$, each spin degree of freedom can
be treated separately.  Still one has to consider for each spin
projection up to $N$ electrons so that the relevant Hilbert space has the
dimension $2^N$.  Therefore, it is more efficient to consider the
single-particle density matrix
\begin{equation}
   R_{\alpha \beta}(t)
   = \langle c_{\alpha s}^\dagger (t) c_{\beta s}(t) \rangle_{t}
   = R^*_{\beta\alpha}(t)= \sum_k \exponent{- \i k \Omega t}
     R_{\alpha\beta,k},
\end{equation}
which is of dimension $N^2$ and nevertheless contains all relevant
information.  The Fourier decomposition in the last expression uses
the fact that, at asymptotically large times, $R_{\alpha\beta}(t)$
becomes time-periodic.

We express the time derivative of $R_{\alpha\beta}(t)$ with the master
equation \eqref{mastereq-gen} and insert for $\tilde
c_\alpha(t,t-\tau)$ the Floquet
representation \eqref{c-t-t'}.  After some algebra, we obtain for the
Fourier coefficients $R_{\alpha\beta,k}$ the equation
\begin{equation}
\label{mastereq}
\begin{alignedat}{1}
    \i(\epsilon_\alpha -\epsilon_\beta +k\hbar & \Omega)R_{\alpha\beta,k}
    \\
     = \frac{\Gamma_\L}{2} \sum_{k'}
    \Big( & \sum_{\beta',k''}
    \langle \varphi_{\beta,k'+k''} | 1 \rangle
    \langle 1 | \varphi_{\beta',k+k''} \rangle
    R_{\alpha\beta',k'}\\
    +&\sum_{\alpha',k''}
    \langle \varphi_{\alpha',k'+k''} | 1 \rangle
    \langle 1 | \varphi_{\alpha,k+k''} \rangle
    R_{\alpha'\beta,k'}\\
    -& \langle \varphi_{\beta,k'-k} | 1 \rangle
    \langle 1 | \varphi_{\alpha,k'} \rangle
    f_\L(\epsilon_{\alpha,k'})\\
    -&
    \langle \varphi_{\beta,k'} | 1 \rangle
    \langle 1 | \varphi_{\alpha,k'+k} \rangle
    f_\L(\epsilon_{\beta,k'}) 
    \Big)\vphantom{\sum_k'}\\
    & \hspace*{-10ex} + \text{terms with the replacement\ }
    \text{$\L, 1 \rightarrow \R, N$}.
\end{alignedat}
\end{equation}
In contrast to the master equation \eqref{eq:masterfourier} for the limit
of strong Coulomb repulsion, no blocking factors $1-f_\mathrm{L/R}$ emerge.
This is characteristic for the non-interacting limit and is also found
in the exact scattering formula \cite{Kohler2005a}.

Using the single-particle density matrix $R_{\alpha\beta}$, one can
evaluate the current expectation value \eqref{current-general}.
Thereby the spin degree of freedom enters simply as a factor $2$ so
that we obtain for the current per spin the expression
\begin{equation}
  \begin{split}
    I =
    \frac{e\Gamma_\L}{\hbar}
    \sum_{\alpha,k}
      \Big[
      &
      \Re\sum_{\beta,k'}
      \langle
      \varphi_{\beta,k'+k}|1\rangle\langle1|\varphi_{\alpha,k}\rangle
      R_{\alpha\beta,k'}
  \\
  & - |\langle 1|\varphi_{\alpha,k}\rangle|^2\,
      f_\L(\epsilon_{\alpha,k})
  \Big] .
  \end{split}
  \label{Imastereq}
\end{equation}
For a detailed derivation see Ref.~\cite{Lehmann2003b}.

\section{Interplay of dipole radiation and Coulomb repulsion}
\label{sec:applications}

In our model Hamiltonian \eqref{eq:full_hamiltonian} we have already
specified the interaction, the lead Hamiltonian, and the wire-lead
coupling.  By contrast, for the Hamiltonian of the driven wire, we
have thus far only assumed that the external field is periodic in
time.  In the following we focus on models where the single-particle
dynamics is determined by the $N$-site tight-binding Hamiltonian
\begin{equation}
\label{dipole}\
\begin{split}
H_{nn'}(t)
= & -\Delta \sum_{n=1}^{N-1}(|n\rangle\langle n+1| +|n+1\rangle\langle n|)
  \\
  & + \sum_{n=1}^N \{E_n + A x_n \cos(\Omega t)\} |n\rangle\langle n| .
\end{split}
\end{equation}
Neighbouring sites are coupled by a tunnel matrix element $\Delta$.
The onsite energies $E_n$ are modulated by a harmonically
time-dependent dipole force, where the amplitude $A$ is given by the
electrical field amplitude multiplied by the electron charge and the
distance between neighbouring sites.  $x_n = \frac{1}{2}(N+1-2n)$
denotes the scaled position of site $|n\rangle$.
Depending on the onsite energies $E_n$, one observes various phenomena
which we discuss in the following.

\subsection{Resonant excitations of bridged molecular wires}
\label{sec:reso}

A frequently studied model is the so-called bridged molecular wire
\cite{Nitzan2001a} sketched in Fig.~\ref{fig:wire_model}. It
consists of a molecule with $N$ sites, where the first and
the last site --- the donor $\ket{1}$ and the acceptor $\ket{N}$
--- are connected to respective leads. The energies
of these two sites are assumed to be close to the chemical
potentials of the respective
leads, $\mu_\L \lesssim E_1 = E_N \lesssim \mu_\R$.  The
remaining $N-2$ orbitals lie well above the chemical potentials at an
energy $E_\mathrm{B}\gg\Delta$ which defines the bridge height.  In
the presence of laser excitations, we expect an enhanced current
whenever the energy quanta of the laser field $\hbar\Omega$ matches the
energy difference between donor/acceptor and one of the $N-2$ bridge
levels.  This photon-assisted tunnelling indeed exhibits resonance peaks
which obey the scaling law $I_\mathrm{peak}\propto A^2/(N-1)\Gamma$
\cite{Kohler2005a}.

\begin{figure}
  \centerline{
  \includegraphics{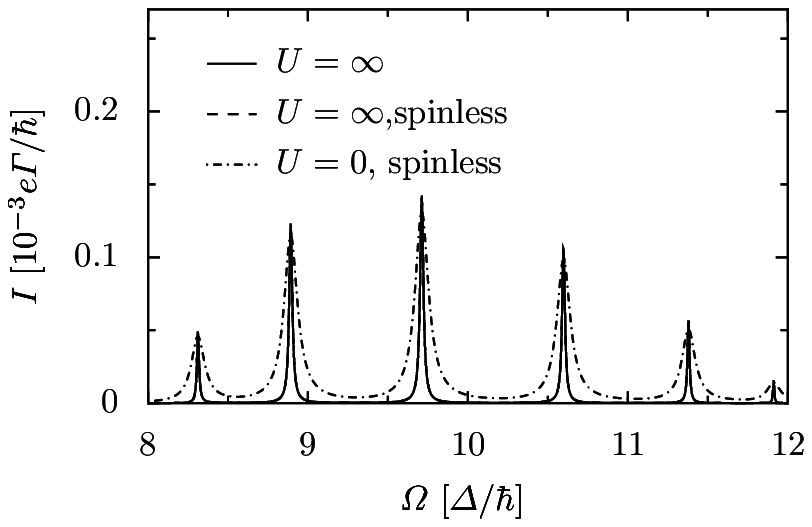}}
  \caption{\label{fig:physrep_414_b_comparison}
  Photon-assisted current in a bridged molecular wire with length
  $N=8$ and height $E_\mathrm{B}=10\Delta$.  Leads with a chemical
  potential difference $eV=5\Delta$ couple to the wire with an effective
  strength $\Gamma=0.1\Delta$.  The driving amplitude is $A=0.3\Delta$.
  For $U=\infty$, the current for spinless electrons coincides with
  the one for real electrons.
  }
\end{figure}%
Figure~\ref{fig:physrep_414_b_comparison} shows the dc current as a
function of the laser frequency for the three approaches considered
herein.  In all these cases, the resonance peaks are at the same
frequencies.  As a further interesting feature, we find that in the
strongly interacting case $U=\infty$, the current does not depend on
whether one includes the spin degree of freedom.  This behaviour can
be understood qualitatively in the following way:
If there is no excess electron on the wire, an electron can
enter from both the ``spin-up channel'' and the ``spin-down channel''.
By contrast, when leaving the wire, the electron spin is preserved so
that only either channel is available. This means that including the
electron spin effectively modifies the in-tunnelling rates by a factor
2. The electron dynamics within the wire, however, is described by the
coherent first term of the master equation \eqref{mastereq-gen} which is
spin-independent.  Consequently, we expect that the spin only plays a
minor role whenever the tunnelling from the donor to the acceptor
becomes the bottleneck for the electrons.  This is indeed the case for
the transport across a barrier considered here.

As compared to the non-interacting case, one notices that strong Coulomb
repulsion modifies the shape of the peaks: They become
slightly higher and much sharper.  This effect is more pronounced for large
wires.  A closer look at the resonance peaks for wires with up to $N=10$
sites (not shown) indicates that the resonance widths scale roughly
like $1/N$.  This is possibly caused by the fact that the Coulomb
repulsion reduces the number of available (many-particle) wire states
and, thus, the number of decay channels. This relates to the
observation made in Ref.~\cite{Kaiser2006a}, namely that Coulomb
repulsion can improve quantum coherence and thereby enhance the
current.

\begin{figure}
  \centerline{
  \includegraphics{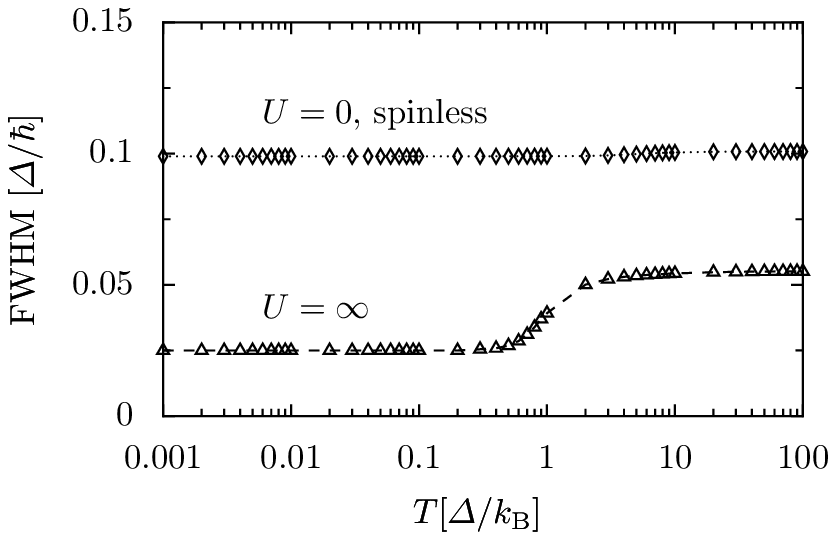}}
  \caption{\label{fig:temp}
  Temperature dependence of the widths of the resonance peaks at
  $\Omega\approx 9.71\Delta/\hbar$.
  The graph shows the full widths at half of the maximum.
  All other parameters are as in Fig.~\ref{fig:physrep_414_b_comparison}.
  For $U=\infty$, the values for spinless electrons and real electrons
  coincide.
  }
\end{figure}%
Since quantum coherence is also temperature dependent, it is natural
to ask whether the resonance peaks become sharper for lower
temperatures. Figure~\ref{fig:temp} shows the width of the central
peak of Fig.~\ref{fig:physrep_414_b_comparison} as a function of the
temperature. While in the non-interacting case, the peak widths are
essentially temperature independent, the situation changes for strong
Coulomb repulsion: There one finds that with an increasing
temperature, the peaks become roughly twice as broad once the
temperature exceeds $T=\Delta/k_\mathrm{B}$.  We attribute this
behaviour to the reduced coherence for thermal energies that are
larger than the tunnelling matrix element.

\subsection{Non-adiabatic electron pumping}
\label{sec:pump}

Another well studied phenomenon in driven transport is coherent
electron pumping, i.e., the creation of a non-vanishing dc current by
ac fields in the absence of any net bias.  For adiabatically slow
driving, this effect exists only in the absence of time-reversal
symmetry \cite{Brouwer1998a, Altshuler1999a, Switkes1999a}.  Beyond
the adiabatic regime, this is no longer the case: For fast,
time-periodic driving fields, it can be shown that the relevant
symmetry is the so-called generalized parity which is defined as the
invariance under spatial
reflection in combination with a time shift by half a driving period
\cite{Kohler2005a}.  Non-adiabatic electron pumping is particularly
interesting because at internal resonances of the central system the
pump current can assume rather large values \cite{Stafford1996a,
Brune1997a, vanderWiel2003a}, while at the same time the current noise is
remarkably low \cite{Strass2005b}.
\begin{figure}
\centerline{\includegraphics{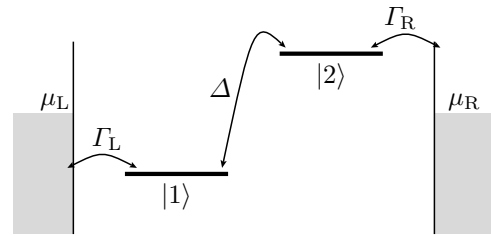}}
\caption{\label{fig:pump-setup}
  Tight-binding model for two coupled quantum dots in pump
  configuration, i.e.\ in the absence of a bias voltage but with
  an internal bias $E_2-E_1\neq 0$ which breaks reflection symmetry.}
\end{figure}%
\begin{figure}
\centerline{\includegraphics{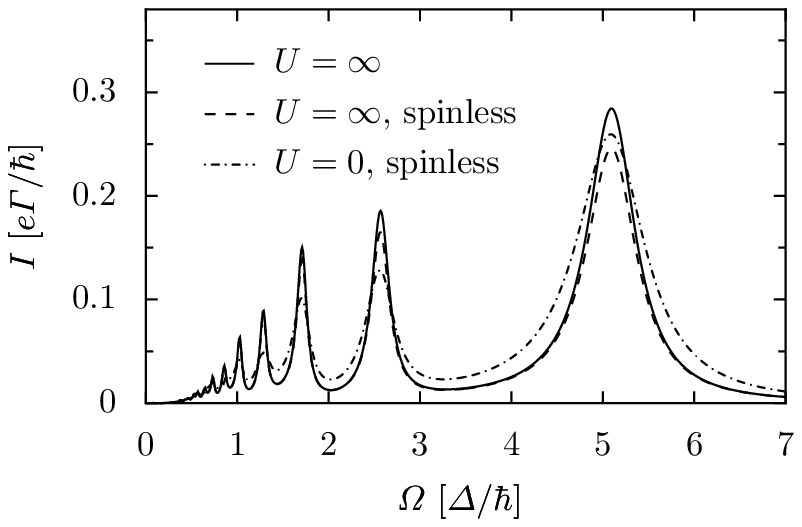}}
\caption{\label{fig:strass_prl}
Pump current for strong Coulomb repulsion and for non-interacting
electrons as a function of the frequency.  The dot levels with
energies $E_{\mathrm{1}}=-2.5\Delta$ and $E_{\mathrm{2}}=2.5\Delta$
are couple to the leads with strength $\Gamma=0.3\Delta$.  The driving
amplitude is $A=3.7\Delta$ and the temperature $k_\mathrm{B}T=0.005\Delta$.}
\end{figure}%

For studying the influence of strong Coulomb repulsion on non-adiabatic
electron pumping, we consider the setup sketched in
Fig.~\ref{fig:pump-setup}.  Of particular interest is the
parameter regime with large internal bias and
intermediate dot-lead coupling because in this regime, the current-to-noise
ratio is most favourable~\cite{Strass2005b}.
The currents obtained for the three considered approaches are shown in
Fig.~\ref{fig:strass_prl}.  Again we find that the spin degree of
freedom is not of major influence, which indicates that the
transport is governed by internal excitations; cf.\ the discussion in
the preceeding subsection.

The influence of the Coulomb repulsion is a modification of the
current peak height up to 5\%.  This means that interactions are here much less
important than for photon-assisted transport:  The reason for this is
that for our pump configuration, one energy level lies below the Fermi
energy while the other lies well above.  Thus in equilibrium for a
sufficiently small dot-lead coupling, the left site is occupied while
the right site is empty, whatever the interaction strength.
Thus, the double dot is populated with only one electron so that
interactions become irrelevant.  Unless the driving amplitude is huge,
this occupation is altered only slightly.  Consequently interactions
do not modify the current significantly.
We emphasise that for strong dot-lead coupling $\Gamma$ and finite interaction
$U$, these arguments no longer hold true.

\subsection{Coherent current control}
\label{sec:cdt}

An intriguing example of quantum control is the so-called coherent
destruction of tunnelling in a double-well potential by a suitable
driving field \cite{Grossmann1991a}, which can be explained
within a rotating-wave approximation: In the
driving dominated regime, the tunnel matrix element is essentially
replaced by an effective tunnel matrix element $\Delta_\mathrm{eff} =
J_0(A/\hbar\Omega)$, where $J_0$ is the zeroth order Bessel function
of the first kind and $A$ the driving amplitude \cite{Grossmann1991b,
Grossmann1992a}.  Related effects have been predicted for driven
tight-binding lattices \cite{Holthaus1992b}.  For driven transport
between two leads, the corresponding situation has been investigated
only recently:  It has been found that driving fields that suppress
tunnelling in a closed driven system, also suppress the current through
the corresponding open system \cite{Lehmann2003a, Kohler2004a}.
\begin{figure}
\centerline{\includegraphics{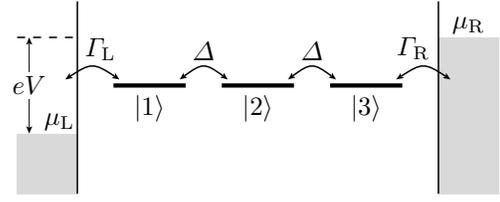}}
\caption{\label{fig:control-setup}
  Triple quantum dot configuration for coherent current control:
  A large bias voltage and the unbiased dot levels with $E_n=0$ ensure optimal
  transport in the absence of the driving.}
\end{figure}%
\label{sec:control}
\begin{figure}
  \centerline{\includegraphics{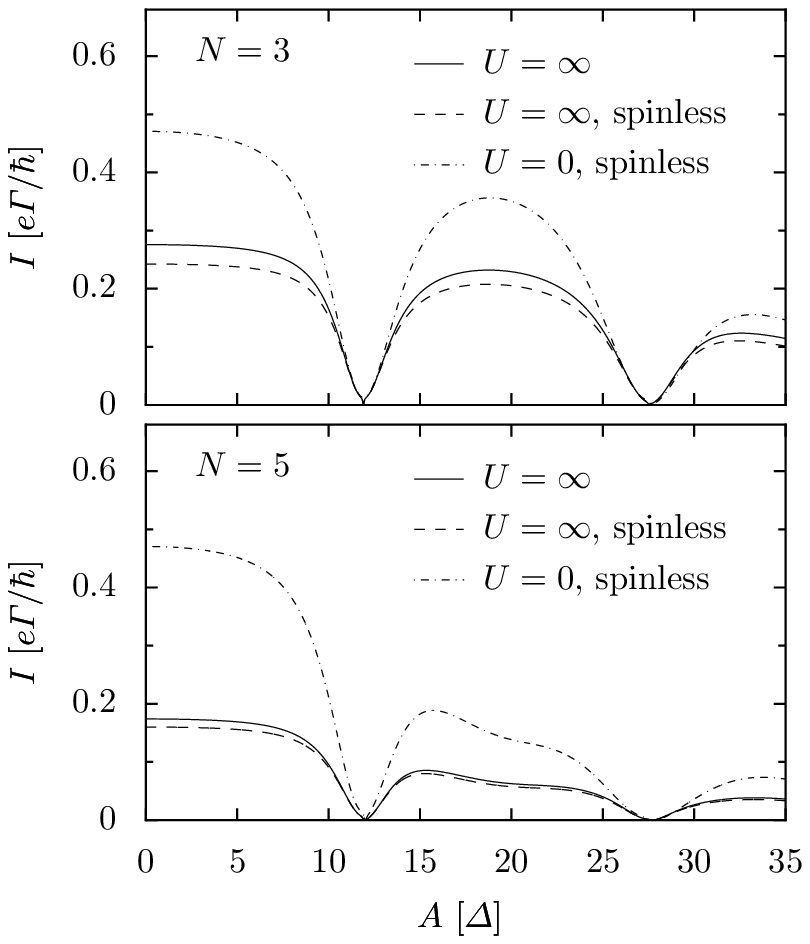}}
  \caption{Coherent current suppression as a function of the driving
  amplitude $A$ for a wire that consists of $N=3$ sites.  The applied
  voltage is $eV=50\Delta$, the driving frequency $\Omega=5\Delta$, and
  the wire-lead coupling $\Gamma=0.5\Delta$.
  \label{fig:cdot}}
\end{figure}%

A setup in which this coherent current suppression can be studied is
sketched in Fig.~\ref{fig:control-setup}.  It is characterized by the
fact that all internal levels lie within the voltage window, i.e.\
below one chemical potential and above the other.  For this system, strong
interaction has already a significant influence on the current
in the absence of any driving field \cite{Kaiser2006a}:  For $U=0$,
the system is half filled, which means it is populated by $N/2$
electron, and the current is independent of $N$.  By contrast for
$U=\infty$, the stationary population is $N/(N+1)$ and the current becomes
$I\propto 1/(N+1)$.

Figure~\ref{fig:cdot} shows the influence of Coulomb repulsion on
the current suppression studied in Ref.~\cite{Camalet2003a}.
Independent of the interaction, one finds that the current
almost vanishes whenever the ratio $A/\hbar\Omega$ matches a zero of
the Bessel function $J_0$.
If the driving amplitude is far from the values for which the current
is suppressed, we observe the behaviour found for the static situation,
namely that Coulomb repulsion reduces the current by a factor
$1/(N+1)$ for spinless electrons \cite{Kaiser2006a}.
If one considers the spin, this factor becomes $2/(2N+1)$.
In the vicinity of the current suppressions, by
contrast, the influence of both the inclusion of the spin and the
interaction is less pronounced.  In this regime, the effective tunnel
matrix element $\Delta_\mathrm{eff}$ is small, so that
tunnelling along the wire happens at a low rate.
This again indicates that whenever the transport is limited by the
dynamics within the wire, the influence of interaction is rather small.

\section{Conclusions}

We studied the influence of strong interaction on the transport
properties of ac-driven coherent conductors.  In particular, we
compared the strongly interacting case with the opposite extreme of
non-interacting electrons.  Moreover, we worked out the relevance of
the spin degree of freedom for weak wire-lead coupling.
In our studies, we considered three archetypical effects, namely
photon-assisted tunnelling through bridged molecular wires,
non-adiabatic electron pumping, and coherent current suppression.

The most significant effect is found for photon-assisted tunnelling
where Coulomb repulsion renders the resonance linewidths much sharper.
Thus unfortunately, interactions might contribute to the difficulties in
photon-assited tunnelling experiments with molecular wires.
By contrast, Coulomb repulsion is not too relevant for electron
pumping in double quantum dots.  For coherent current suppression, the
same holds true only for parameters for which the current is already
significantly reduced.  Outside this region, one finds that Coulomb
repulsion reduces the current essentially in the same way as in the
absence of driving.

The two extreme cases of zero and very strong interaction do not
necessarily allow a simple interpolation.  Thus, it is desireable to
extend the present studies to finite values of the interaction strength,
which requires the generalisation of our formalism to at least a
second excess electron.  Moreover, the dc current is certainly not the
only relevant quantity for the characterisation of the electron
transport.  Investigating the influence of Coulomb repulsion on, e.g., the
current noise would complement the picture drawn above.

\begin{acknowledgement}

We thank M. Strass and A. Nitzan for interesting
discussions.  This work has been supported by Deutsche
Forschungsgemeinschaft through SFB 484 and SPP 1243.  One of us (FJK)
acknowledges funding by Bayerisches Staatsministerium f\"ur Wissenschaft,
Forschung und Kunst through Elitenetzwerk Bayern.

\end{acknowledgement}

\end{document}